\documentstyle[epsf,twocolumn]{article}

\catcode`@=11
\newdimen\p@ \p@=1pt 
\newdimen\z@ \z@=0pt 
\newskip\z@skip \z@skip=0pt plus0pt minus0pt

\def\gne{\mathrel{\mathpalette\@vesim>}}
\def\lne{\mathrel{\mathpalette\@vesim<}} 
\def\@vesim#1#2{\lower4\p@\vbox{\baselineskip\z@skip\lineskip.5\p@
\ialign{$\m@th#1\hfil##\hfil$\crcr#2\crcr\sim\crcr}}}

\begin{document}
\title{
	\bf
	An Extended Anderson Model that shows 
	Decreasing Resistivity with Decreasing Temperature
}

\author{
	S. Suzuki$^1$, O. Sakai and Y. Shimizu$^a$ \\
	Department of Physics, Tohoku University, Sendai 980-77,				 Japan \\
	$^a$ Department of Applied Physics, Tohoku University, 
		Sendai 980-77, Japan
}

\date{
	\normalsize
	\bf
	submitted to Solid State Communication \\
	received 16 May 1997; accepted 19 June 1997 by T. Tsuzuki
}

\maketitle
\begin{abstract}
   The resistivity for an extended two channel Anderson model
   is calculated.
   When the temperature decreases, it decreases logarithmically,
   and has $T^{1/2}$ anomaly at very low temperatures,
   as seen in some dilute U alloys.
   The single-particle and the spin excitation spectra are
   also calculated.
   The low energy properties and the temperature dependence 
   of the resistivity in the low temperature region 
   are well described by the energy scale deduced from 
   the zero-energy limit of the spin excitation spectrum.
\end{abstract}
\footnotetext[1]{E-mail: suzukis@cmpt01.phys.tohoku.ac.jp}
\linebreak
\linebreak
\linebreak
The two channel Kondo model (TCKM) has been studied extensively
since Nozi\`{e}res and Blandin pointed out the possibility
of the non-Fermi liquid (NFL) behaviours of its ground state~[1].
The analytic solutions for the thermodynamic quantities have been
obtained by means of the Bethe-Ansatz method~[2].
The $\gamma$-coefficient and the susceptibility, 
$\chi$, diverge as $-$ln$T$ at low temperatures.

Cox has noted that a model for the
electronic state of ions with $f^2$-configuration,
such as ${\rm U^{4+}}$, can be mapped on the TCKM
when the crystalline field ground state is
the non-Kramers doublet~[3].
Many studies have been reported on
the applicability of the TCKM to real systems [4-8].

In experiments, a number of dilute U-ion alloys
have been known to show the NFL behaviours.
Amitsuka and Sakakibara have shown that $\gamma$ and $\chi$ of 
${\rm Th_{1-x}U_xRu_2Si_2}$ 
have the $-$ln$T$ divergence in the dilute limit.
At the same time, the resistivity decreases logarithmically
with decreasing temperature, 
and then decreases in the form $R(0)-c_RT^{1/2}$
with negative $c_R$ at very low temperatures~[9].

Usually $c_R$ is expected to be positive for the TCKM.
Though Affleck {\it et al.} have noted that $c_R$ can be negative
in the strong exchange cases~[10],
recent studies
based on the numerical renormalization group (NRG) method 
have shown that the coefficients of the logarithmic 
divergence of $\gamma$ and $\chi$ are very small 
for such model~[7,11].
Therefore it seems inconsistent with the experimental results.

In a previous paper, we have presented an extended Anderson model
which has the possibility of showing the divergent susceptibility
and decreasing resistivity
with decreasing temperature~[12].
The single particle excitation spectrum $\rho_f(\omega)$ is
given as $\rho_f(\omega)\cong \rho_f(0)-c_f|\omega|^{1/2}$
in the $\omega\rightarrow 0$ limit.
The coefficient $c_f$ is negative in the weak exchange cases
and positive in the strong exchange cases
contrasted to the TCKM.
Therefore, the resistivity is expected to decrease
with decreasing temperature in the weak exchange cases.
In this letter, we calculate the temperature dependence of
the resistivity for the model, 
and show that it actually decreases at low temperatures.
In addition, we show that
the temperature dependence of the resistivity, 
the spin excitation spectrum, $\rho_s(\omega)$,
and the single particle excitation spectrum, $\rho_f(\omega)$,
are well described by one energy scale in the low energy region.
The energy scale is extracted from the low energy limit 
of the spin excitation, $\rho_s(0)$.

The extended Anderson model we consider is,
\begin{eqnarray}
H & = & H_{c}+H_{c-f}+H_{f}, \\
H_{f} & = & \!\sum_{m=(0,\pm1),\alpha=\pm1}\!
	\varepsilon_{fm}n_{m\alpha} \\
& + & \frac{U}{2}\sum_{((m=\pm1)\alpha)\neq((m'=\pm1)\alpha')}
	n_{m\alpha}n_{m'\alpha'} \\
& + & J\!\sum_{(m=\pm1)\{\alpha\}}\!
	\vec{\sigma}_{\alpha_1\alpha^{'}_1}
	\vec{\sigma}_{\alpha_2\alpha^{'}_2}
	f_{m\alpha_1}^{\dagger}f_{m\alpha^{'}_1}
	f_{0\alpha_2}^{\dagger}f_{0\alpha^{'}_2}, \\
H_{c-f} & = & \sum_{k}\sum_{(m=\pm1)\alpha}
	V(f^{\dagger}_{m\alpha} c_{km\alpha}+h.c.), \\
H_{c} & = & \sum_{k}\sum_{(m=\pm1)\alpha}
	\varepsilon_{k}c_{km\alpha}^{\dagger}c_{km\alpha}.
\end{eqnarray}
The operator $f_{m\alpha} (c_{km\alpha})$ is the annihilation
operator of the localized $f$-electron (conduction electron
with wave number $k$).
The channel is denoted by $m$, which takes 0 and $\pm1$,
and each has spin freedom, $\alpha=\pm1$.
The quantities $U,J$ and $\varepsilon_k$ are the Coulomb
interaction constant, the exchange constant and
the conduction band energy, respectively.
We assume that the band has width from $-D$ to $D$
with $D=1$.
The density of states of the conduction electrons, $\rho$, is
chosen as constant $\rho=1/2D$.
The energy level of the 0-channel, $\varepsilon_{f0}$ 
is assumed to be deep enough so that $n_{f0}$ 
is always restricted to 1.
In addition, we assume that the energy levels and hybridization
matrices of other two channels are independent on $m$, i.e.
$\varepsilon_{fm}=\varepsilon_f$ and $V_m=V$ for $m=\pm1$.
If the orbits of the 0-channel and the $\pm1$ channels are
ascribed, respectively, to the doublet $\Gamma_7$ orbit
and the quartet $\Gamma_8$ orbit in the cubic crystalline field,
the present model is similar to the original Anderson model
of Cox, from which he derived the TCKM based on 
the Schrieffer-Wolff transformation and 
restricting the effective manifold to 
$f^2$-configuration~[13].

In this study, we calculate the electric resistivity, $R(T)$,
neglecting the vertex correction term.
The normalized resistivity is given by
\begin{eqnarray}
R(T)/R(0) & = & \frac{L_{01}(0)}{L_{01}(T)},
\end{eqnarray}
where $L_{ml}$ is defined as
\begin{equation}
L_{ml}(T)=\int^{+\infty}_{-\infty} \left(-\frac{\partial f}
{\partial \omega}\right) \omega^{m} \tau(\omega,T)^{l} d\omega.
\end{equation}
Here, the relaxation time for the conduction electrons, 
$\tau(\omega,T)$,
is given by using the single particle excitation spectrum of
$f$-electron $\rho_f(\omega,T)$ as
\begin{equation}
\tau(\omega,T)=\frac{1}{(\pi V^2\rho)\rho_f(\omega,T)}.
\label{tau-eq}
\end{equation}

The excitation spectra at finite temperatures are calculated by 
using the NRG~[14,15].
Parameters are chosen mostly to be the same as in
our previous paper~[12].
The spin excitation spectrum, $\rho_s(\omega)$, 
is defined as the imaginary part of the susceptibility of 
the operartor, $s^z_f=\sum_{m=0,\pm1}s^z_{fm}$.
The channel excitation spectrum, $\rho_c(\omega)$, is that of 
$n^c_f=\sum_{m=\pm1,\alpha=\pm1}mn_{fm\alpha}$.
We define an energy scale, $T^0_{\rm K}$,
which corresponds to the Kondo temperature of
the fictitious two channel Anderson model without the 0-channel spin
(FTCAM)[15,16].
\begin{figure}
\epsfxsize=8cm \epsfbox{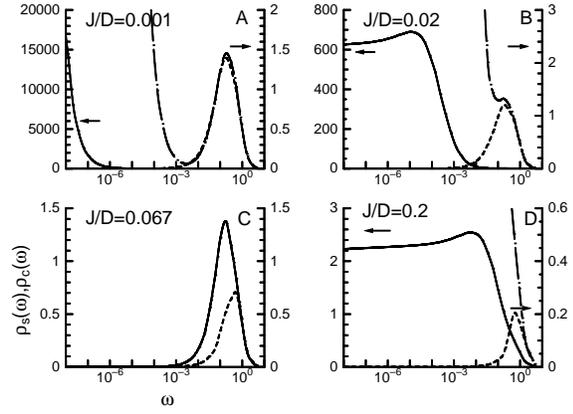}
\caption{
	The spin and channel excitation at $T=0$ 
	as a function of the logarithm of the energy,
	for various $J$. 
	$U=1.0$, $\varepsilon_f=-1.5$ and $V^2\rho=0.1$.
	The left ordinate is the scale for 
	the solid line which shows $\rho_s(\omega)$.
	The right ordinate is the scale for 
	the dashed and dot-dashed lines which show 
	$\rho_s(\omega)$ and $\rho_c(\omega)$ in the
	magnified scale.
	The Kondo temperature, $T^0_{\rm K}$, for 
	the fictitious two channel Anderson model is 
	estimated to be $\approx$ 0.19.
	The sign of the coefficient for the $|\omega|^{1/2}$
	term in the single particle excitation, $c_f$, is 
	negative for A (less than $-10^4$) and B($-33$),
	almost zero for C ($-0.3$),
	and positive for D ($0.65$).
}
\label{raw_mag}
\end{figure}

Figure 1 shows 
$\rho_s(\omega)$ and 
$\rho_c(\omega)$ for various $J$ 
from the weak exchange cases, $J/T^0_{\rm K}<0.35$,
to the strong exchange cases, $J/T^0_{\rm K}>0.35$.
In the weak exchange cases (Figs. 1A and B),
$\rho_s(\omega)$ has two different structures:
one at very low energy and another at about $T^0_{\rm K}$.
Moreover, $\rho_s(\omega)$ has the a finite value 
in the $\omega\rightarrow 0$ limit, in Fig. 1B.
It shows the plateau and decreases steeply at the energy
characterized by $0.01/\rho_s(0)\sim 10^{-5}$ as $\omega$ increases.
The finiteness of $\rho_s(0)$ is contrasted to 
the Fermi liquid behaviour going to zero proportionally to $\omega$
in the $\omega\rightarrow0$ limit, and it indicates 
the $-$ln$T$ divergence of the spin susceptibility~[17].
As the exchange coupling, $J$, decreases, 
this NFL structure moves to the lower energy region 
and $\rho_s(0)$ increases (Fig. 1A).
In the strong exchange cases (Fig. 1D),
$\rho_s(0)\times T^0_{\rm K}$ seems to be always less than 
1.25~[12].
In Fig. 1A, there is a distinct peak 
at $\omega\sim T^0_{\rm K}(=0.19)$.
In Fig. 1B, the peak merges into the tail 
of the NFL structure, and looks like a shoulder.
In the weak exchange cases, the height of 
this peak is much smaller than that of 
the NFL structure.
When $J$ is small enough so that the peak is isolated,
it is quite similar to the peak in $\rho_c(\omega)$ 
(Fig. 1A).
These peaks in $\rho_s(\omega)$ and $\rho_c(\omega)$
are formed by the Kondo effect of the FTCAM~[16].

\begin{figure}
\epsfxsize=8cm \epsfbox{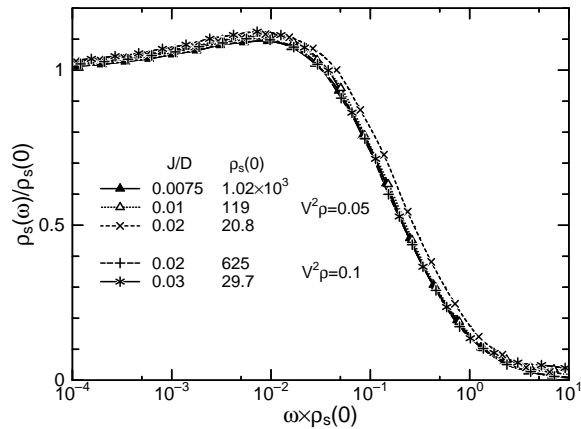}
\caption{
	The normalized spin excitation spectra as a function
	of the normalized energy $\omega\times\rho_s(0)$, 
	for various $J$ and $V$. $U=1.0$ and $\varepsilon_f=-1.5$,
	other parameters are shown in the figure.
	For the all cases, the sign of $c_f$ is negative.
}
\label{norm_mag}
\end{figure}
In Fig. 2, the normalized spin excitation spectra,
$\rho_s(\omega)/\rho_s(0)$, are shown as a function of 
$\omega\rho_s(0)$.
The spectra show good overlapping in the cases of large $\rho_s(0)$.
When $\rho_s(0)$ is large enough so that the NFL structure is
well separated from the $\rho_c$-like peak at $\omega\sim T^0_{\rm K}$,
the NFL structure seems to be characterized 
by one energy scale deduced by $\rho_s(0)$.

\begin{figure}
\epsfxsize=8cm \epsfbox{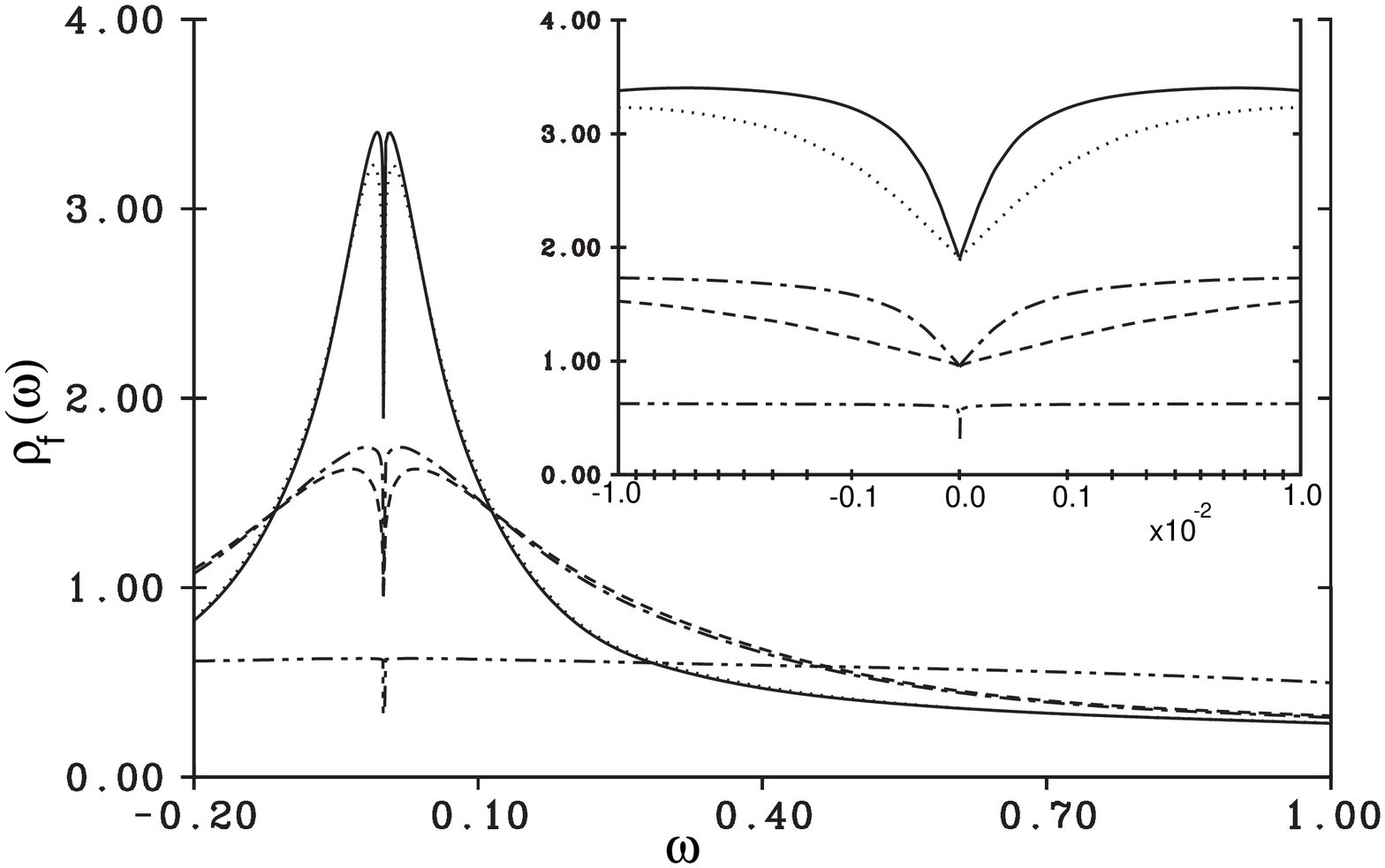}
\caption{
	The single particle excitation spectra as a function of
	energy at $T=0$, for various parameters.
	The main figure shows the spectrum in the energy range 
	$\approx T^0_{\rm K}$.
	The inset figure shows the cusp structure at around 
	the Fermi energy.
	The abscissa of the inset is in the square-root 
	of the energy.
	The parameters are: 
	solid line
		($U=1.0$, $J=0.0075$, $V^2\rho=0.05$,
		 $T^0_{\rm K}=0.11$,
		 $\rho_s(0)=1.02\times 10^3$),
	dotted line
		(1.0, 0.01, 0.05, 0.11, 119),
	dotted-dashed line
		(1.0, 0.02, 0.1, 0.19, 625),
	dashed line
		(1.0, 0.03, 0.1, 0.19, 29.7),
	two-dotted-dashed line
		(0.1, 0.04, 0.3, 1.04, $2.50\times 10^7$).
}
\label{spe_raw}
\end{figure}
The single particle excitation spectra, $\rho_f(\omega)$, at
$T=0$ are shown in Fig.~3.
When $\omega$ decreases, 
$\rho_f(\omega)$ increases in the energy region 
$\omega\gne T^0_{\rm K}$ almost analogously 
to the spectrum for the FTCAM.
It shows the peaks in both sides of the Fermi energy,
and decreases to half of the intensity for the FTCAM
as $\omega$ goes to zero~[12].
In the very low energy region,
$\rho_f(\omega)$ has the cusp singularity which is described as 
$\rho_f(\omega)\cong \rho_f(0)-c_f|\omega|^{1/2}$.
As $J$ decreases,
the width of the cusp structure becomes narrow.

\begin{figure}
\epsfxsize=8cm \epsfbox{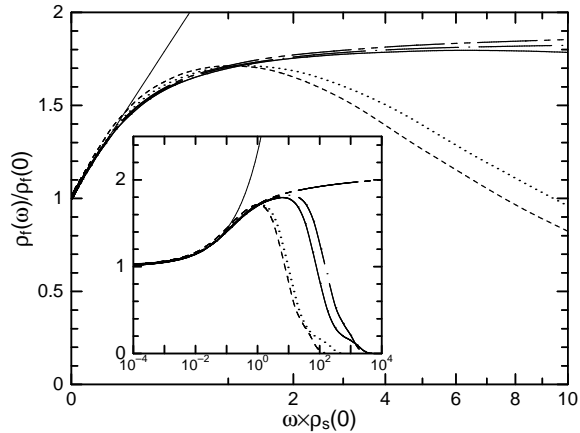}
\caption{
	The normalized single particle excitation as 
	a function of the normalized energy $\omega\rho_s(0)$.
	The main figure shows $\rho_f(\omega)$,
	the abscissa is in the square-root 
	of the energy.
	The straight fine line represents a fit to 
	$\rho_f(\omega)=\rho_f(0)-c_f|\omega|^{1/2}$.
	The inset figure shows same quantities 
	in the logarithmic energy scale.
	The parameters for each lines are same as those in Fig.~3.
}
\label{spe_norm}
\end{figure}
Figure 4 shows the normalized single particle excitation,
$\rho_f(\omega)/\rho_f(0)$ at $T=0$,
as a function of $\omega\rho_s(0)$.
The low energy part seems to have a universal shape 
when we use $1/\rho_s(0)$ as the energy unit.
In the energy region $\omega<0.1/\rho_s(0)$,
$\rho_f(\omega)$ decreases with decreasing $\omega$ almost
linearly with $|\omega|^{1/2}$.
Moreover, the universality seems to be valid 
in the higher energy region 
where the curves deviate from the simple $|\omega|^{1/2}$ law.

\begin{figure}
\epsfxsize=8cm \epsfbox{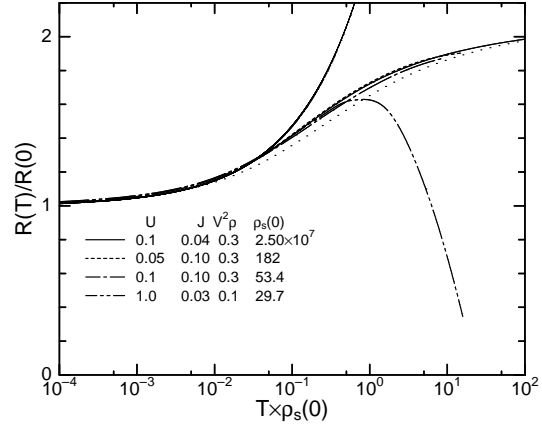}
\caption{
	The normalized resistivity as a function of normalized
	temperature $T\rho_s(0)$.
	The parameters are:
	solid line
		($U=0.1$, $J=0.04$, $V^2\rho=0.3$,
		$\rho_s(0)=2.50\times 10^7$), 
	dashed line	
		(0.5, 0.10, 0.3, 182),
	one-dotted-dashed line
		(0.1, 0.10, 0.3, 53.4),
	two-dotted-dashed line
		(1.0, 0.03, 0.1, 29.7).
	The fine line represents a fit to $R(T)=R(0)-c_RT^{1/2}$.
	The dotted line is the fictitious resistivity 
	which is calculated without the temperature 
	dependence of $\rho_f(\omega)$.
	The parameters are (0.1, 0.04, 0.3, $2.50\times 10^7$),
	same as those of the solid line.
}
\label{res_norm}
\end{figure}
The electric resistivity is shown in Fig. 5.
As expected from the numerical results of $\rho_f(\omega)$,
the resistivity decreases with decreasing temperature.
The temperature dependence of the decreasing behaviour 
is well described by using $1/\rho_s(0)$ as the temperature unit.
However, we note that
the resistivity increases with decreasing temperature 
at higher temperatures $T\gg T^0_{\rm K}$,
similarly to the resistivity of the usual Kondo effect.
At low temperatures $T<0.05/\rho_s(0)$,
the resistivity is given as $R(T)\cong R(0)-c_RT^{1/2}$,
with negative $c_R$.
Affleck {\it et al.} have predicted the relations between 
the coefficient for the $T^{1/2}$ term and 
that for the $-$ln$T$ divergent term in thermodynamic 
quantities for the TCKM~[10].
These relations will be checked for the present model 
in the future paper.
In the middle temperature region, $0.05\leq T\rho_s(0)\leq 1$,
$R(T)$ seems to have the logarithmic temperature dependence.

The fictitious resistivity for which 
we use $\rho_f(\omega,T=0)$ instead of $\rho_f(\omega,T)$ 
in eq. (1) is shown in Fig. 5.
Though it is slightly smaller than the proper resistivity,
its temperature dependence is not different much from
that of the proper resistivity~[15].

\begin{figure}
\epsfxsize=4cm \epsfbox{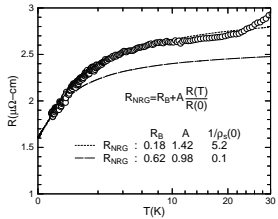}
\caption{
	The comparison of the resistivity between 
	our numerical result and the experimental 
	result.
	The circles show the experimental value for 
	${\rm Th_{0.97}U_{0.03}Ru_2Si_2}$.
	The fitting function is $R_B+A(R(T)/R(0))$.
	The parameter $R_B$ is ascribed as 
	the residual resistivity constant of
	the ${\rm ThRu_2Si_2}$.
	The dashed line is given by fitting resistivity at 10K,
	and the dotted-dashed line is given by supposing 
	$2.6\mu\Omega{\rm cm}$ of the plateau 
	as the unitarity value plus $R_B$.
	In each case, $1/\rho_s(0)$ is determined by fitting
	the $T^{1/2}$ term at very low temperatures.
}
\label{vs_expr}
\end{figure}
Figure 6 compares our numerical result
of the resistivity with that of
${\rm Th_{1-x}U_xRu_2Si_2}$~[9].
The calculated result is qualitatively
similar to the experimental result.
But the ambiguities for determination of the parameters 
remain~[18].
Further study is needed to give a quantitative 
conclusion~[19].
The present model has been by no means directly
related to the realistic model of U ion,
but the calculation of the model will support the scenario
to explain the anomalous properties based on
the NFL behaviours of the TCKM type.

In summary, we have calculated the dynamical excitations and
the electric resistivity for an extended two channel Anderson model.
The leading temperature dependence of the resistivity 
is proportional to $T^{1/2}$ at very low temperatures.
Also there is the temperature region 
where the resistivity decreases logarithmically.
The behaviours are qualitatively similar to 
the experimental results for ${\rm Th_{1-x}U_xRu_2Si_2}$.
We have shown that 
the low energy properties are well arranged by 
one energy scale deduced from the quantity $\rho_s(0)$,
when the normalized exchange coupling is small, $J/T^0_{\rm K}\ll 0.35$.
As $J/T^0_{\rm K}$ increases,
the energy region characterized by 
the two channel Kondo effect
spreads to the higher energy side.
The sign changing of $c_f$ seems to occur 
when the spread of the TCKM region 
becomes comparable to $T^0_{\rm K}$.

In this letter, we reported only the symmetric case
which the occupation number is $n_f=3$ accounting the local spin.
Even for the small occupation number cases, $n_f<3$, 
the $T$ dependence of the resistivity
at low temperatures is essentially equivalent to
that of the symmetric case.
Detail will be shown in the future paper.

The authors would like to thank 
T. Sakakibara, K. Kuwahara,
R. Takayama, M. Koga, S. Takagi and H. Suzuki 
for helpful discussions.
The experimental results of the resistivity for 
${\rm Th_{1-x}U_xRu_2Si_2}$ is courtesy by H. Amitsuka.
This work was partly supported by Grant-in-Aids No. 06244104,
No. 09640451 and No. 09244202 from the Ministry of Education, 
Science and Culture of Japan.
The numerical computation was partly performed at the 
Computer Center of Institute for Molecular Science 
(Okazaki National Research Institute), 
the Computer Center of Tohoku University, 
and the Supercomputer Center of Institute for Solid State Physics 
(University of Tokyo).


\begin{thebibliography}{1}
\bibitem{ferro}
   We note that the atomic ferromagnetic exchange interaction
   between $f$-electrons are mapped to the anisotropic
   antiferromagnetic coupling between $\Gamma_7$ and $\Gamma_8$
   electrons.
\bibitem{FTCAM}
   The FTCAM is the Anderson model with four fold degeneracy
   that the usual Fermi liquid picture is applied.
   In this model,
   the channel excitation is equal to the magnetic excitation,
   and its peak position is good 
   approximation of the usual definition of the Kondo 
   temperature, see reference 15.
\bibitem{future1}
   The $-{\rm ln}T$ anomaly is also found in the spin susceptibility
   which is calculated directly by using the NRG.
   The numerical results will be published in the near future.
\bibitem{fitting}
   In the present case, $R(0)$ is half of the unitarity limit
   of the high temperatures.
   Hence, 
   from the $T=0$ limit ($R_B+A$) and 
   the high temperature limit ($R_B+2A$)
   of the experimental resistivity,
   we can estimate $R_B$ and $A$.
   The lines in Fig.~6 reaches the unitarity limit
   at very high temperature $T\gne 10^3$.
   The value at the unitarity limit of the dotted-dashed line is 
   $2.6\mu\Omega{\rm cm}$,
   and that of the dotted line is $3.0\mu\Omega{\rm cm}$.
\end{thebibliography}
\end{document}